\documentclass[fleqn,12pt,twoside]{article}
\usepackage{amsbsy}
\usepackage{amssymb}
\usepackage[headings]{espcrc1}

\readRCS
$Id: espcrc1.tex,v 1.2 2004/02/24 11:22:11 spepping Exp $
\usepackage{graphicx}
\usepackage{epsfig}
\usepackage[figuresright]{rotating}

\newcommand{\be}{\begin{equation}}
\newcommand{\ee}{\end{equation}}
\newcommand{\bra}{\langle}
\newcommand{\ket}{\rangle}
\newcommand{\id}{{1\!\!1}}
\newcommand{\om}{\omega}
\newcommand{\nv}{{\mathbf n}}
\newcommand{\pv}{{\mathbf p}}

\newcommand{\vecx}{{\mathbf x}}
\newcommand{\thetav}{\boldsymbol\theta}
\newcommand{\vecnul}{{\mathbf 0}}

\newcommand{\AmS}{{\protect\the\textfont2
  A\kern-.1667em\lower.5ex\hbox{M}\kern-.125emS}}

\title{Meson spectral functions at nonzero momentum in hot 
QCD\footnote{talk presented at Strong \& Electroweak Matter 
(SEWM2006), BNL, United States, May 10-13 2006}
}

\author{Gert Aarts\address[UWS]{Department of Physics, 
	Swansea University, Swansea, United Kingdom}\thanks{speaker, 
PPARC Advanced Fellow},
        Chris Allton\addressmark,
        Justin Foley\addressmark,
        Simon Hands\addressmark\, and
        Seyong Kim\address{Department of Physics, Sejong University, 
	Seoul, Korea}
}
       
\runtitle{Meson spectral functions at nonzero momentum in hot QCD}
\runauthor{Gert Aarts et.\ al.}

\begin{document}

\maketitle

\begin{abstract}

We present preliminary results for meson spectral functions at nonzero 
momentum, obtained from quenched lattice QCD simulations at finite 
temperature using the Maximal Entropy Method.
 Twisted boundary conditions are used to have access to many momenta 
$p\sim T$. 
 For light quarks, we observe a drastic modification when heating the 
system from below to above $T_c$. In particular, for the vector spectral 
density we find a nonzero spectral weight at all energies.

\end{abstract}

\vspace{0.5cm}

Spectral functions provide insight into the real-time dynamics of the 
quark-gluon plasma. In the context of heavy-ion collisions and RHIC, 
lattice QCD studies carried out so far address, among others, the survival 
of charmonium up to $T\sim 1.6T_c$ 
\cite{Asakawa:2003re,Datta:2003ww,Umeda:2002vr,Morrin:2005zq}, the rate of 
dilepton production \cite{Karsch:2001uw}, and transport coefficients 
\cite{Karsch:1986cq,Aarts:2002cc,Gupta:2003zh,Nakamura:2004sy,Petreczky:2005nh}. 
Recently, analytical studies of spectral functions in strongly coupled 
gauge theories have also become available, using the gauge-gravity duality 
conjecture for ${\cal N}=4$ supersymmetric Yang-Mills theory
 \cite{Teaney:2006nc,Kovtun:2006pf}.

Meson spectral functions $\rho_H(\om,\pv)$ can be obtained from euclidean 
correlators obtained in lattice simulations by inverting the relation
 \be
 G_H(\tau,\pv) = \int_0^\infty \frac{d\om}{2\pi}\, K(\tau,\om) 
\rho_H(\om,\pv), 
\;\;\;\;\;\;\;\;\;\;\;\;
K(\tau,\om) = \frac{\cosh[\om(\tau-1/2T)]}{\sinh(\om/2T)},
\label{eq1}
\ee
 using the Maximal Entropy Method (MEM) \cite{Asakawa:2000tr}. Here 
$G_H(\tau,\vecx)=\bra J_H(\tau,\vecx)J_H^\dagger(0,\vecnul)\ket$ and 
$J_H(\tau,\vecx)=\bar q(\tau,\vecx) \Gamma_H q(\tau,\vecx)$ with 
$\Gamma_H=\{\id,\gamma_5, \gamma_\mu, \gamma_5\gamma_\mu\}$.  Usually only 
zero momentum ($\pv=\vecnul$) is considered. In order to better understand 
the resulting spectral functions, especially at high temperature, it may 
be beneficial to consider spectral functions at nonzero momentum as well. 
There are several reasons for this \cite{Aarts:2005hg}. At high 
temperature, spectral functions are no longer expected to be sharply 
peaked. Instead one expects them to have a rich structure and contain 
e.g.\ a contribution below the lightcone ($\om<p=|\pv|$) and a 
momentum-dependent threshold above the lightcone. For very soft momentum 
and energy ($\om\sim p \ll T$) one expects hydrodynamical structure in 
spectral functions related to conserved currents (see refs.\ 
\cite{Aarts:2005hg,Karsch:2003wy} for more details on both continuum and 
lattice meson spectral functions at infinite temperature; for other recent 
weak-coupling calculations, see e.g.\ ref.\ \cite{Alberico:2006wc}). 
Finding specific momentum-dependent features in spectral functions 
obtained via MEM may help in the interpretation.

On a lattice with $N_\sigma$ sites in a spatial and $N_\tau$ sites in the 
temporal direction, the smallest nonzero momentum in units of the 
temperature is $p/T = 2\pi N_\tau/N_\sigma$. To have access to the 
hydrodynamic regime with many momenta $p/T\lesssim 1$ at high temperature 
(with large enough $N_\tau$ for MEM purposes) is extremely demanding. For 
that reason we have adopted so-called twisted boundary conditions, 
following closely the formulation given in ref.\ \cite{Flynn:2005in}. 
Combining the twisting with a standard spatial Fourier transformation, the 
meson momentum reads $\pv L = 2\pi \nv - (\thetav_1 - \thetav_2)$, where 
$\nv$ are integers, $\thetav_1$ and $\thetav_2$ are the twist angles of 
the two quarks, and $L=aN_\sigma$. In this contribution we present results 
on two lattices, below and above $T_c \approx 270$ MeV, specified by

{\flushleft
 \begin{tabular}{llllll}
  cold: & $48^3\times 24$, & $\beta=6.5$, & $a^{-1}\sim 4$ GeV, 
 & $T\sim 160$ MeV, &100 propagators, \\
  hot: & $64^3\times 24$, & $\beta=7.192$, & $a^{-1}\sim 10$ GeV, 
 & $T\sim 420$ MeV, &100 propagators. 
 \end{tabular}
}

\vspace*{0.3cm}
\noindent 
 The estimates for the lattice spacing and the temperature are taken from 
ref.\ \cite{Datta:2003ww}. Note that $N_\tau$ is kept fixed, whereas the 
other parameters vary. Combining four different twist angles in a number 
of ways, we have $\sim 20$ different momenta in the range $0<p/T<4$ on the 
hot lattice. The results shown here are for light staggered quarks, 
$am=0.01$ ($m/T=0.24$).

\begin{figure}[t]
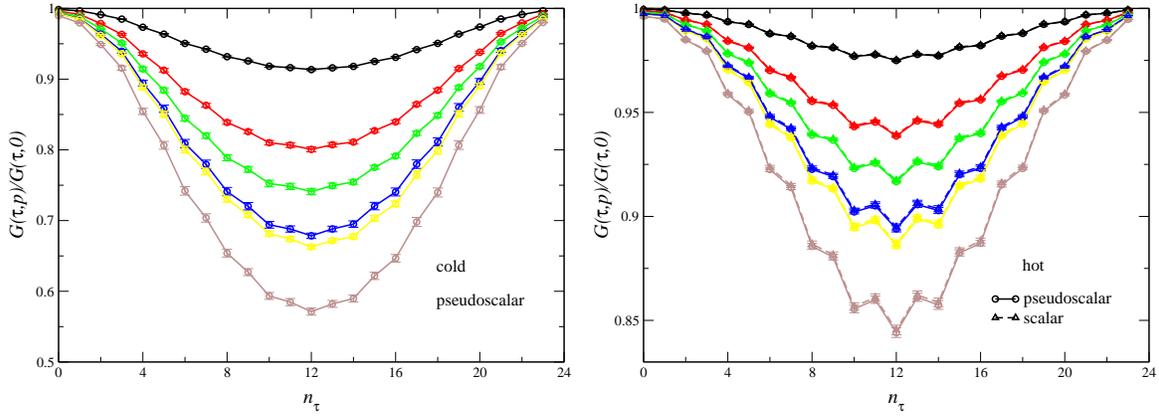

\centerline{
	\epsfig{figure=Gratio_PSS_b6.500_m0.01_nconf.100m.eps,width=7.5cm}
	\hspace*{0.cm}
	\epsfig{figure=Gratio_PSS_b7.192_m0.01_nconf.100m.eps,width=7.5cm}
}
\vspace*{-0.6cm}
 \caption{Euclidean correlator at nonzero momentum normalized with the 
correlator at zero momentum, $G(\tau, \pv)/G(\tau,\vecnul)$, in the 
pseudoscalar channel below $T_c$ (left) and in the scalar and pseudoscalar 
channel above $T_c$ (right). Because of chiral symmetry restoration, the 
latter are hardly distinguishable. The momenta are $pL = 2.0$, $3.14$, 
$3.72$, $4.25$, $4.36$, $5.2$ (top to bottom).
 }
 \label{figG}
\end{figure}

In fig.\ \ref{figG} the pseudoscalar correlators both in the cold and the 
hot phase are presented. In order to see the relative effect of increasing 
the momentum we show the ratio of $G(\tau,\pv)$ and $G(\tau,\vecnul)$. 
While in the cold phase the relative effect is large (up to 45\% for the 
largest momentum shown), in the hot phase the relative effect is much less 
(up to 15\% for the largest momentum shown). In fact, in the hot phase 
this effect is of the same order as the one obtained in a free quark 
calculation at infinite temperature (note that in the ratio the operator 
renormalization factors cancel). In the hot phase we also present the 
scalar correlator to demonstrate chiral symmetry restoration above $T_c$.

\begin{figure}[t]
\centerline{
	\epsfig{figure=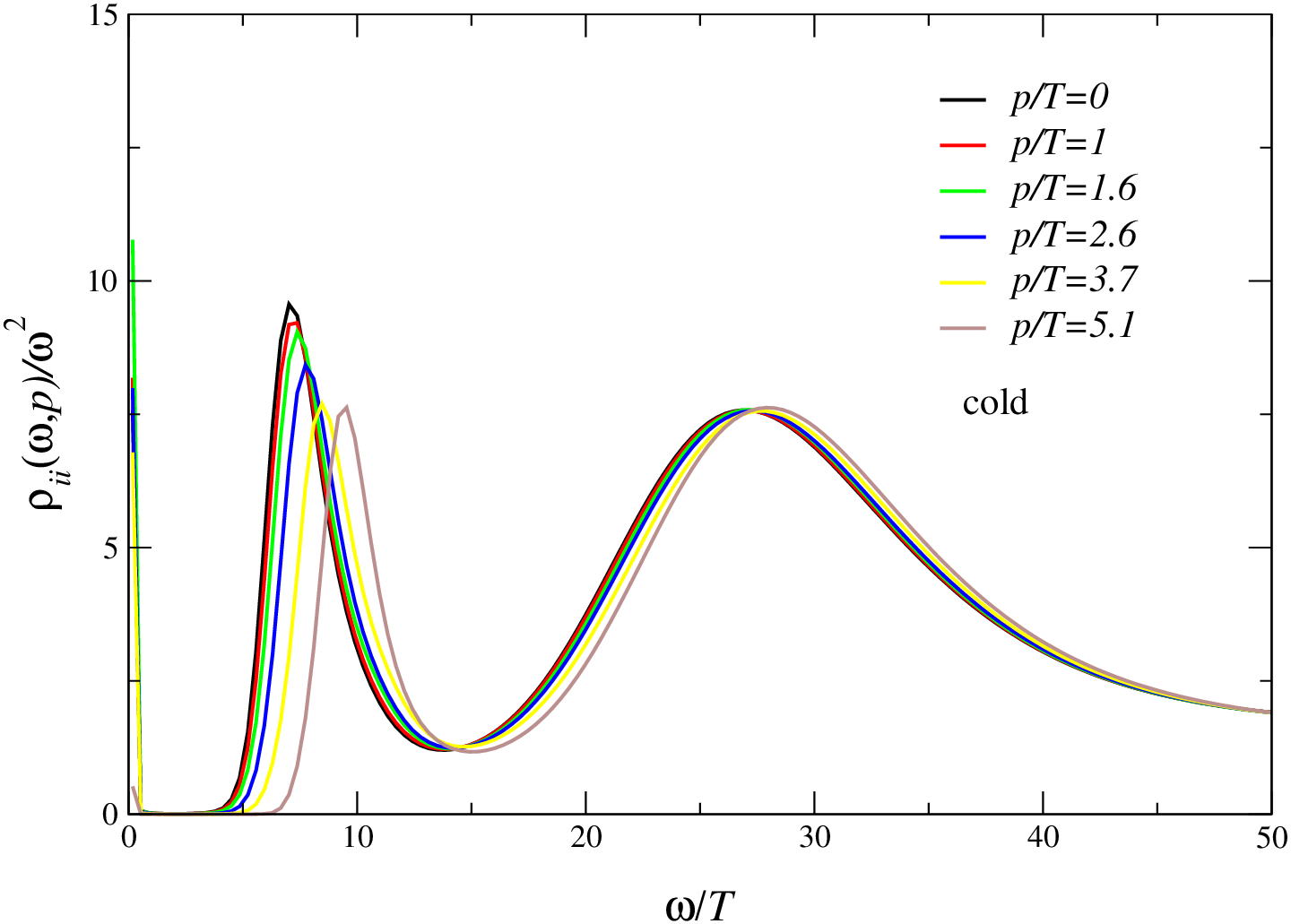,width=7.5cm}
	\hspace*{0.cm}
	\epsfig{figure=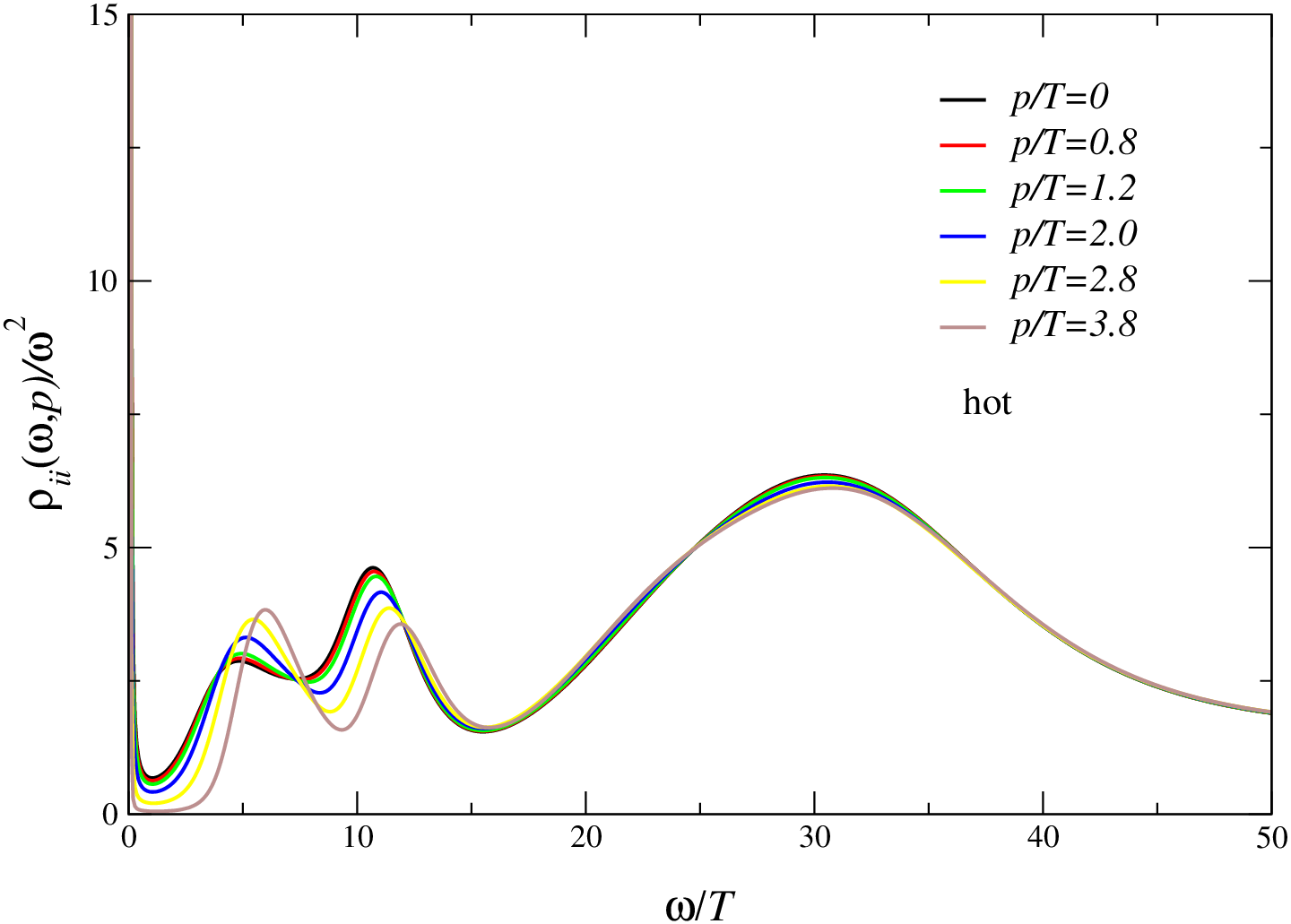,width=7.5cm}
}
\vspace*{-0.6cm}
 \caption{Spectral functions in the vector channel 
$\rho_{ii}(\om,\pv)/\om^2$ (summed over $i=1,2,3$) for various values of 
$p$ below $T_c$ (left) and above $T_c$ (right).
 }
 \label{figrho1}
\vspace*{0.6cm}
\centerline{
	\epsfig{figure=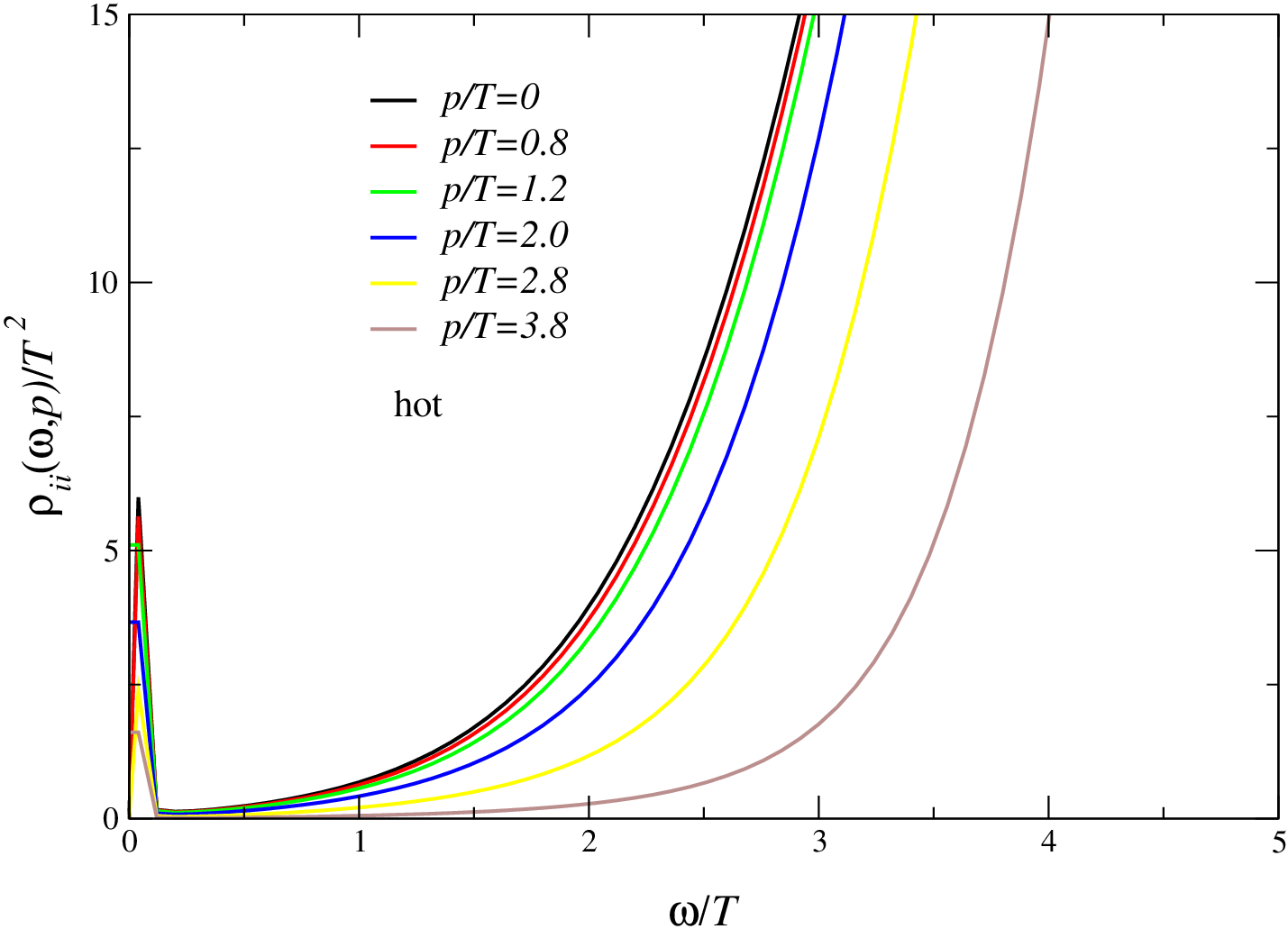,width=7.5cm}
	\hspace*{0.cm}
	\epsfig{figure=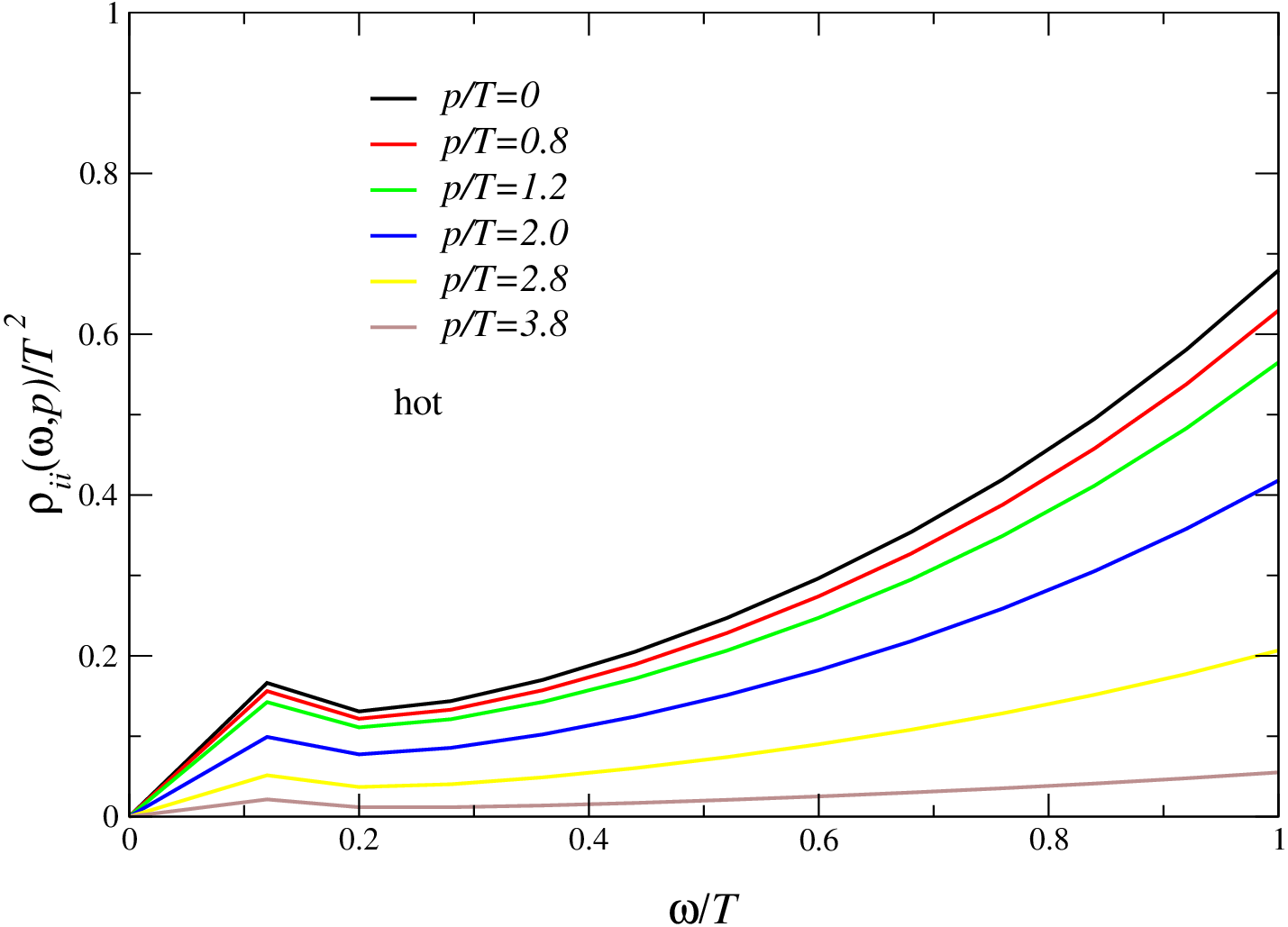,width=7.5cm}
}
\vspace*{-0.6cm}
 \caption{Left: vector spectral functions above $T_c$, 
$\rho_{ii}(\om,\pv)/T^2$, for various values of $p$. The sharp peak near 
$\om=0$ is an artefact of the MEM analysis. 
 Right: blow-up of the small-energy region with the sharp peak near zero 
removed. 
 }
 \label{figrho2}
\end{figure}

The MEM analysis for staggered fermions is more complicated than for 
Wilson fermions due to the mixing between two signals in the correlators. 
The equivalent of relation (\ref{eq1}) for staggered fermions reads
 \be
 G_H^{\rm stag}(\tau,\pv) = 2\int_0^\infty \frac{d\om}{2\pi}\, K(\tau,\om) 
\left[ 
\rho_H(\om,\pv) - (-1)^{\tau/a}\tilde \rho_H(\om,\pv) \right], 
\label{eq2}
\ee
 where $\tilde \rho_H$ is related to $\rho_H$ by changing $\Gamma_H$ to 
$\tilde\Gamma_H = \gamma_4\gamma_5\Gamma_H$. In practice we perform an 
independent MEM analysis on the even and odd time slices, obtaining 
$\rho_{\rm even} = \rho_H-\tilde \rho_H$ and $\rho_{\rm odd} = 
\rho_H+\tilde \rho_H$ and combine these to get $\rho_H$. 

In fig.\ \ref{figrho1} we show spectral functions in the vector channel 
($\Gamma_H=\gamma_i$, summed over $i=1,2,3$), normalized with $\om^2$, in 
the cold and the hot phase. In the cold phase there is a clear bound state 
peak whose position shifts to larger energy with increasing momentum. The 
second broad structure at $\om/T \sim 30$ ($a\om\sim 1.25$) is probably a 
lattice artefact, due to the finite Brillouin zone \cite{Aarts:2005hg}. In 
the hot phase the bound state appears to have ``melted'', although some 
structure remains. This requires further study.
 Note that the horizontal axis covers a huge range $0<\om/T<50$. To focus 
on the region relevant for thermal physics, we show a blow-up of 
$\rho_{ii}$, now normalized with $T^2$, in fig.\ \ref{figrho2}. A sudden 
increase is visible, which, especially for the larger momenta, appears to 
occur roughly when $\om\sim p$. This is reminiscent of the threshold 
behaviour observed at weak coupling. The sharp peak at $\om\sim0$ is an 
artefact of the MEM analysis and depends e.g. on the resolution along the 
$\om$ axis. Removing this peak by hand and zooming in even further yields 
the result shown in fig.\ \ref{figrho2} (right). This region is of special 
interest for hydrodynamics and transport, since the electrical 
conductivity can be defined from the slope of this spectral function at 
vanishing energy via the Kubo formula, $\sigma = \lim_{\om\to 
0}\rho_{ii}(\om,\vecnul)/(6\om)$. We note that the spectral weight is 
nonzero for all $\om$, which is expected at high temperature. Moreover, 
for the smallest momenta there is a hint of a bump at small $\om$. Whether 
this indeed corresponds to the expected hydrodynamical structure, and in 
particular whether the electrical conductivity can be (reliably 
\cite{Aarts:2002cc}) extracted from the slope, requires further analysis 
\cite{inprep}.

\end{document}